\documentclass[twocolumn]{aastex631}
\usepackage{CJK}
\usepackage{amsmath}
\shorttitle{Cosmologically-coupled mass growth of BHs \& mass gap problem}
\shortauthors{Gao \& Li}

\begin{document}
\begin{CJK*}{UTF8}{gbsn}

\title{Can cosmologically-coupled mass growth of black holes solve the mass gap problem?}

\author[0000-0002-0822-0337]{Shi-Jie, Gao (高世杰)}\email{gaosj@smail.nju.edu.cn}
\affiliation{School of Astronomy and Space Science, Nanjing University, Nanjing 210023, P. R. China}
\affiliation{Key Laboratory of Modern Astronomy and Astrophysics, Ministry of Education (Nanjing University), Nanjing 210023, P. R. China}
\author[0000-0002-0584-8145]{Xiang-Dong, Li (李向东)}\email{lixd@nju.edu.cn}
\affiliation{School of Astronomy and Space Science, Nanjing University, Nanjing 210023, P. R. China}
\affiliation{Key Laboratory of Modern Astronomy and Astrophysics, Ministry of Education (Nanjing University), Nanjing 210023, P. R. China}

\begin{abstract}
Observations of elliptical galaxies suggest that black holes (BHs) might serve as dark energy candidates, coupled to the expansion of the Universe. According to this hypothesis, the mass of a BH could increase as the Universe expands. BH low-mass X-ray binaries (LMXBs) in the Galactic disk were born several gigayears ago, making the coupling effect potentially significant. In this work, we calculate the evolution of BH binaries with a binary population synthesis method to examine the possible influence of cosmologically-coupled growth of BHs, if it really exists. The measured masses of the compact objects in LMXBs show a gap around $\sim 2.5-5~{\rm M_\odot}$, separating the most massive neutron stars from the least massive BHs. Our calculated results indicate that, considering the mass growth seem to (partially) account for the mass gap and the formation of compact BH LMXBs, alleviating the challenges in modeling the formation and evolution of BH LMXBs with traditional theory. However, critical observational evidence like the detection of intermediate-mass black hole binaries is required to test this hypothesis.
\end{abstract}


\section{Introduction}

\begin{table*}[!ht]
    \centering
    \caption{Types, BH masses ($M_{\rm BH}$), donor masses ($M_2$) and orbital periods ($P_{\rm orb}$) of dynamically confirmed intermediate-mass X-ray binaries (IMXBs)/LMXBs.\label{tab:obs}}
    \setlength{\tabcolsep}{5pt}
    \begin{tabular}{llllll}
    \hline
    Name&Type&$M_{\rm BH}~[\rm M_\odot]$&$M_2~[{\rm M_\odot}]$&$P_{\rm orb}~[{\rm day}]$&References\\
    \hline
    GRS~1915+105&LMXB&$12.4_{-1.8}^{+2.0}$&$0.52\pm0.31$&$33.85$&\cite{Reid+2014}\\
    V404~Cyg&LMXB&$9.0_{-0.6}^{+0.2}$&$0.54\pm0.05$&$6.473$&\cite{Khargharia+2010}\\
    V4641~Sgr&IMXB&$6.4\pm0.6$&$2.9\pm0.4$&$2.8173$&\cite{MacDonald+2014}\\
    4U~1543--47&IMXB&$9.4\pm1.0$&$2.7\pm1$&$2.7$&\cite{Shafee+2006}\\
    GRO~J1655--40&LMXB&$5.31\pm0.07$&$1.45\pm0.35$&$2.62168$&\cite{Beer+2002}\\
    GS~1354--64&LMXB&$>7.6\pm0.7$&$>0.988$&$2.54451$&\cite{Casares+2009}\\
    GX~339--4&LMXB&$9_{-1.2}^{+1.6}$&$0.7\pm0.4$&$1.75$&\cite{Parker+2016}\\
    LMC~X--3&IMXB&$6.98\pm0.56$&$3.63\pm0.57$&$1.7048089$&\cite{Orosz+2014}\\
    XTE~J1550--564&LMXB&$9.10\pm0.61$&$0.30\pm0.07$&$1.5420333$&\cite{Orosz+2011}\\
    MAXI~J1820+070&LMXB&$8.48_{-0.72}^{+0.79}$&$0.61_{-0.12}^{+0.13}$&$0.6903$&\cite{Torres+2020}\\
    H1705--250&LMXB&$6.4\pm1.5$&$0.25\pm0.17$&$0.5228$&\cite{Remillard+1996,Harlaftis+1997}\\
    GRS~1124--68&LMXB&$11.0_{-1.4}^{+2.1}$&$0.89_{-0.11}^{+0.18}$&$0.43260249$&\cite{Wu+2015,Wu+2016}\\
    MAXI~J1305--704&LMXB&$8.9_{-1.0}^{+1.6}$&$0.43\pm0.16$&$0.394$&\cite{MataSanchez+2021}\\
    4U~1957+11&LMXB&$\sim 4.6$&$\lesssim 1$&$0.39$&\cite{Barillier+2023}\\
    GS~2000+251&LMXB&$5.5-8.5$&$0.16-0.47$&$0.34$&\cite{Charles+1991,Ioannou+2004}\\
    A0620--00&LMXB&$6.61\pm0.25$&$0.40\pm0.045$&$0.3230160$&\cite{Johannsen+2009,Cantrell+2010}\\
    GRS~1009--45&LMXB&$8.5\pm1$&$0.54\pm0.1$&$0.285$&\cite{Filippenko+1999,Macias+2011}\\
    XTE~1859+226&LMXB&$7.85\pm0.46$&$0.55\pm0.16$&$0.276$&\cite{Motta+2022,Yanes-Rizo+2022}\\
    GRO~J0422+32&LMXB&$2.7_{-0.5}^{+0.7}$&$0.33_{-0.20}^{+0.28}$&$0.2121600$&\cite{Casares+2022}\\
    XTE~J1118+480&LMXB&$8.3_{-0.14}^{+0.28}$&$0.22\pm0.07$&$0.16993404$&\cite{Gonzalez+2012}\\
    Swift~J1375--0933&LMXB&$10.9_{-1.6}^{+1.7}$&$0.42_{-0.03}^{+0.04}$&$0.106969$&\cite{Casares+2022}\\
    MAXI~J1659--352&LMXB&$3.3-7.5$&$0.06-0.22$&$0.100$&\cite{Torres+2021}\\
    MAXI~J0637--430&LMXB&$5.1\pm1.6$&$0.25\pm0.07$&$0.092$&\cite{Soria+2022}\\
    \hline
    \end{tabular}
\end{table*}

Observations of stellar-mass black hole (BH) mergers \citep[e.g.,][]{Abbott+2016} and supermassive BHs with the Event Horizon Telescope \citep{EHT+2019}, and BH spin measurements in X-ray binaries \citep[e.g.,][]{Miller-Jones+2021} have shown perfect agreement with the Kerr solution of general relativity. However, the Kerr solution reduces to flat spacetime at spatial infinity, which is inconsistent with the Universe described by the perturbed Friedman-Robertson-Walker cosmology \citep{Planck+2020}. To reconcile the incompatibility, it has been suggested that BHs are singularity-free and have vacuum energy interiors \citep[e.g.,][]{Dymnikova+1992,Dymnikova+2000,Dymnikova+2003,Beltracchi+2019}, which mean that BHs can be coupled to the expansion of the Universe. It leads to an increase of their interior stress-energy (hence the masses of BHs) with the expansion of the Universe \citep{Croker+2019}. The mass ($M_{\rm BH}$) growth of a BH can be described by
\begin{equation}
    M_{\rm BH}(a)=M_{\rm BH}(a_i)\left(\frac{a}{a_i}\right)^k,
\end{equation}
where $a$ is the scale factor of Friedmann-Lema\^itre-Robertson-Walker metric, $a_i$ ($a_i<a$) is the scale factor when BHs become coupled to the expansion of the Universe, and $k=-3w$ is a free parameter describing the coupling strength of an astrophysical object in the interior of which the energy density $\rho$ and the pressure $p$ obey $p=w\rho$ \citep{Croker+2021}. For BHs observed in the Milky Way, this relation reduces to $M_{\rm BH}(z=0)=M_{\rm BH}(z=z_i)(1+z_i)^k$, where $z_i=1/a_i-1$ is the redshift when a BH was formed with $M_{\rm BH}(z=z_i)$.

Recently, \cite{Farrah+2023} found the evidence for cosmologically-coupled mass growth among supermassive BHs in elliptical galaxies over $0\lesssim z\lesssim 2.5$. Their results exclude $k=0$ at 99.98\% confidence and give $k=3.11_{-1.33}^{+1.19}$ at 90\% confidence, consistent with vacuum energy ($w=-1$) interior BH models. This results have captured broad attention, with substantial discussions and comments available in the literature \citep[e.g.,][]{Mistele+2023,Wang+2023,Avelino+2023,Parnovsky+2023,Lei+2023,Andrae+2023,Ghodla+2023}.

BH low-mass X-ray binaries (LMXBs) have low-mass companions ($M_2\lesssim 2~{\rm M_\odot}$) as their mass donors. Dozens of BH LMXBs have been dynamically confirmed (see \autoref{tab:obs}). These binaries, which have ages $>10^{9}~{\rm years}$ \citep{Tauris+2006}, were born in the early Universe ($z\lesssim 2$)\footnote{There is a possibility that some of the BH LMXBs were formed recently by dynamically interactions in dense environment.}. As a result, the influence of cosmologically-coupled growth, if really worked, should have ben significant. Though the formation mechanism of BH LMXBs is not fully understood, it is widely believed that they have evolved from primordial binaries consisting of a massive star and an intermediate/low-mass companion \citep[for a review, see][]{Li+2015}. The conventional formation pathways generally involve a phase of common envelope evolution \citep[CEE,][]{Paczynski+1976,Ivanova+2013} when the primary star (the BH's progenitor) evolves off main-sequence and transfers mass to the low-mass secondary star. The main issue is that the orbital energy of the low-mass star is typically insufficient to expel the massive envelope of the BH's progenitor \citep{Portegies+1997,Kalogera+1999,Kiel+2006,Yungelson+2008}. This would cause merger of the two stars rather with a close binary left.

Neutron stars (NSs) and BHs are the end products of massive stars. A smooth distribution of stellar initial masses may intuitively result in a continuous mass distribution between NSs and BHs. However, there is absence or dearth of observed BHs with masses in the range of $\sim 2.5-5~{\rm M_\odot}$ in LMXBs \citep[for a recent review, see][]{Shao+2022}. The origin of this mass gap is still under debate. The proposed explanations include different growth time-scales of the instabilities driving the explosion of massive stars which affect the amount of the ejected envelope mass \citep{Fryer+2012,Belczynski+2012}, disruption of the binaries with low-mass BHs due to supernova-driven natal kicks \citep{Mandel+2021}, or potential observational artifacts arising from systematic uncertainties in ellipsoidal fits \citep{Kreidberg+2012}.

If BHs can server as vacuum energy with equation of state described by $p=-\rho$, $k$ is equal to $3$, while for NSs, $w\sim 0.05-0.1$ and $-k$ is $\sim0.15-0.3$ \citep{Croker+2019}. Since the cosmological shift for positive $w$ appears as an energy loss rather than mass growth \citep{Croker+2019}, a mass gap could be naturally left between the most massive NSs and the less massive BHs. This cosmological coupling not only leads to mass growth of BHs, but also influences the evolution of the binary orbits and hence the observational characteristics of BH LMXBs. In this work, we perform detailed evolutionary calculations and binary population synthesis to investigate whether cosmological coupling can result in a significant mass gap in LMXBs and its impact on the evolution of BH LMXBs.

This article is organized as follows. In \autoref{sec:method}, we introduce the physical considerations and the binary evolution model. We present our numerically calculated results and compare them with observations in \autoref{sec:res}. Finally, we conclude and discuss our results in \autoref{sec:dis}.

\section{Methods and calculations}\label{sec:method}

\begin{figure*}[!ht]
    \centering
    \includegraphics[width=\linewidth]{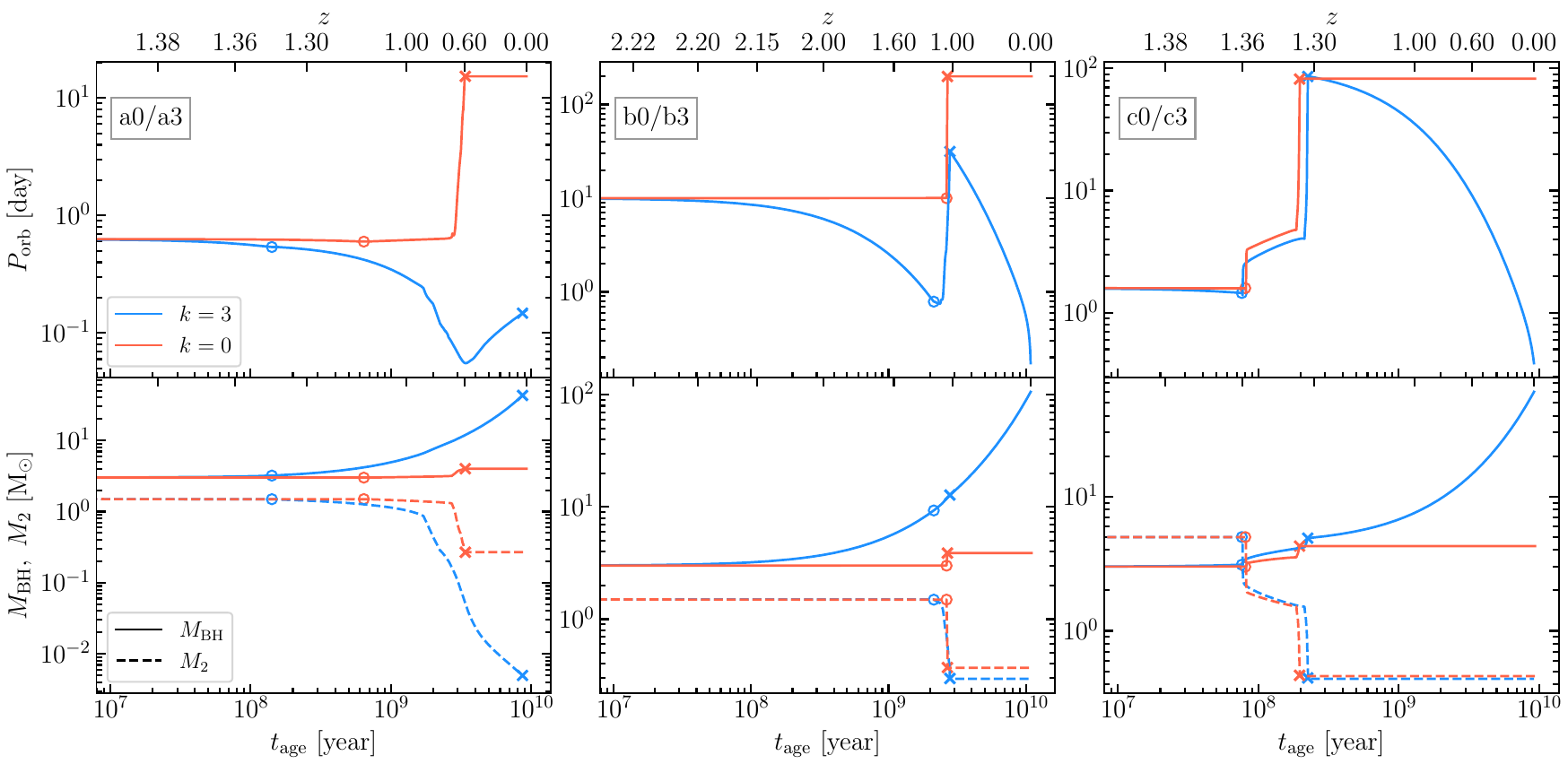}
    \caption{Orbital periods ($P_{\rm orb}$), BH mass ($M_{\rm BH}$) and companion mass ($M_{\rm 2}$) as a function of stellar age ($t_{\rm age}$) and redshift ($z$) for tracks~a0/a3 (left), b0/b3 (middle), and c0/c3 (right). The red and blue lines correspond to the $k=0$ and $k=3$ cases, respectively. The solid and dashed lines in the lower panel show the evolution of the BH and the companion, respectively. The circles and crosses indicate the onset and end of RLOF, respectively.
    \label{fig:mesa}}
\end{figure*}

\begin{table*}[!htb]
    \caption{The parameters of the six evolutionary tracks, including the lookback time $t_{\rm lb}$, redshift $z$ at the birth of the BHs, the initial/final BH masses $M_{\rm BH,i}$/$M_{\rm BH,f}$, companion masses $M_{\rm 2,i}$/$M_{\rm 2,f}$, orbital periods $P_{\rm orb,i}$/$P_{\rm orb,f}$ and mass transfer lifetime $\Delta t_{\rm mt}$. \label{tab:mesaRes}}
    \centering
    \setlength{\tabcolsep}{4pt}
    \begin{tabular}{cccccccccccc}
        \hline
        Tracks&Cases&$t_{\rm lb}~{\rm [Gyr]}$&$z$&$M_{\rm BH,i}~{\rm [M_\odot]}$&$M_{\rm 2,i}~{\rm [M_\odot]}$&$P_{\rm orb,i}~{\rm [day]}$&$M_{\rm BH,f}~{\rm [M_\odot]}$&$M_{\rm 2,f}~{\rm [M_\odot]}$&$P_{\rm orb,f}~{\rm [day]}$&$\Delta t_{\rm mt}~{\rm [year]}$\\
        \hline
        a0&$k=0$&   9.228&   1.388&     3.0&     1.5&    0.63&   4.01&   0.27&  15.31&$ 2.70\times10^9$\\
        a3&$k=3$&   9.228&   1.388&     3.0&     1.5&    0.63&  48.12&   0.004&   0.16&$ 8.54\times10^9$\\
        \hline
        b0&$k=0$&  10.847&   2.230&     3.0&     1.5&   10.00&   3.89&   0.37& 199.65&$ 3.44\times10^7$\\
        b3&$k=3$&  10.847&   2.230&     3.0&     1.5&   10.00& 106.41&   0.29&   0.17&$ 6.70\times10^8$\\
        \hline
        c0&$k=0$&   9.228&   1.388&     3.0&     5.0&    1.58&   4.28&   0.46&  82.51&$ 1.13\times10^8$\\
        c3&$k=3$&   9.228&   1.388&     3.0&     5.0&    1.58&  60.15&   0.44&   0.38&$ 1.45\times10^8$\\
        \hline
    \end{tabular}
\end{table*}

The cosmological coupling introduces no preferred spatial directions. Hence, both angular momentum and eccentricity of the binary are unaffected \citep{Croker+2020}. According to angular momentum conservation in Newtonian binary evolution, the mass growth leads to an extra change \footnote{This change is independent of other mechanisms like magnetic braking, gravitational wave emission and mass transfer process.} (see also Eq.~28 in \citealt{Croker+2020}) in the orbital separation $A$, i.e.,
\begin{equation}\label{eq:sep}
\frac{A(a)}{A(a_i)}=\left[\frac{M_{\rm BH}(a_i)}{M_{\rm BH}(a)}\right]^2\frac{M_2+M_{\rm BH}(a)}{M_2+M_{\rm BH}(a_i)}.
\end{equation}
The quadratic component dominates 
and always leads to decrease in the orbital separation. We apply the cosmological-constant cold dark matter ($\Lambda$CDM) model in our calculation and use the cosmological parameters from the latest cosmic microwave background radiation measurements \citep{Planck+2020}, i.e., $H_0=67.7~{\rm km~s^{-1}~Mpc^{-1}}$, $\Omega_{\rm M}=0.31$, $\Omega_{\Lambda}=0.69$, and the age of the Universe is $13.78~{\rm Gyr}$. We explore two cases $k=0$ and $k=3$, corresponding to without and with cosmologically-coupled growth of BH mass considered, respectively.

\subsection{Example evolutionary calculation}

We use the state-of-the-art stellar evolution code \texttt{MESA}\footnote{The codes to reproduce our simulations are available at \url{https://doi.org/10.5281/zenodo.8167249}.} \citep[version 22.11.1,][]{MESA+2011,MESA+2013,MESA+2015,MESA+2018,MESA+2019,MESA+2023} to evolve binary stars composed of a BH and a zero-age main sequence donor star\footnote{The lifetime of a BH's progenitor star is typically $\lesssim 5~{\rm Myr}$, which can be ignored for the evolution of both a low-mass companion and the Universe ($z<30$).}. The orbital change (\autoref{eq:sep}) due to BH's mass growth is implemented in the code.  We adopt solar metallicity for the stars ($Z=0.02$). To illustrate the effects of cosmologically-coupled growth, we present three example evolutionary tracks (a, b and c) for the cases of $k=0$ (a0, b0 and c0) and $k=3$ (a3, b3 and c3). The initial and final states of these tracks are summarized in \autoref{tab:mesaRes}. Tracks a and b compare the evolution of close and wide LMXBs, and tracks c demonstrates the evolution of intermediate-mass X-ray binaries (IMXBs), which are thought to be the main predecessors of LMXBs.

\autoref{fig:mesa} shows the orbital period, BH mass and companion mass as a function of the stellar age and redshift for tracks~a0/a3 (left), b0/b3 (middle) and c0/c3 (right). The red and blue lines depict the results in the $k=0$ and $k=3$ cases, respectively. The solid and dashed lines in the lower panel represent the evolution of the BH mass and the companion mass,  and the circles and crosses denote the onset and end of Roche-lobe overflow (RLOF), respectively. 

In the left panel, for tracks~a0, the initial orbital period is above the bifurcation period \citep{Pylyser+1988}, so the orbit period increases with the mass transfer to be $15.3~{\rm day}$. The BH mass grows to be $4.01~{\rm M_\odot}$ due to accretion, while the donor star evolves to be a $0.27~{\rm M_\odot}$ white dwarf (WD). However, for track~a3, the BH's mass growth prior to the mass transfer causes the orbit to shrink.  Consequently, the mass transfer starts earlier, and the binary orbit decreases with mass transfer until the mass ratio is reversed. The binary finally consists of a $48.1$~${\rm M_\odot}$ BH and a $0.004$~${\rm M_\odot}$ dwarf star in a $0.16$~${\rm day}$ orbit.

The middle panel shows the evolutionary tracks~b0 and b3 where both binaries eventually evolve into BH+helium WD binaries, but with significantly different final orbital periods. In the case of $k=0$, the orbit expands due to the donor star's expansion to the red giant branch. A $3.9$~${\rm M_\odot}$ BH and a  $0.37$~${\rm M_\odot}$ helium WD are left in a $200$~${\rm day}$ orbit. In the case of $k=3$, the binary contracts before RLOF starts due to the mass growth of BH. Then, the mass transfer causes the binary orbit to expand. Interestingly, after the detachment of the donor, a $0.29$~$\rm M_\odot$ helium WD is formed, and the orbital period decreases dramatically due to the BH's mass growth. The final mass of the BH is $106~{\rm M_\odot}$. The right panel of \autoref{fig:mesa} shows evolutionary tracks~c0 and c3 with a $5~{\rm M_\odot}$ companion star. The evolution shows similar features as in tracks~b0 and b3, except that the companion evolves into a carbon-oxygen WD.

The last column of \autoref{tab:mesaRes} presents the duration of RLOF. It is seen that the mass transfer lifetimes in the case of $k=3$ are longer compared with in the case of $k=0$. This is because in the former case mass transfer starts earlier and ends later. The more extended duration of mass transfer leads to increase in the number of mass-transferring binaries (see below).

\subsection{Binary population synthesis}

\begin{table*}[!ht]
    \caption{Binary population synthesis models and results. Here $N_{\rm tot}$ is the predicted total number of I/LMXBs, $N_{\rm 15}$ and $N_{\rm mg}$ are the numbers of I/LMXBs with BH mass in the range of $<15~{\rm M_\odot}$ and $<5~{\rm M_\odot}$ (mass gap), respectively.}
    \centering
    \setlength{\tabcolsep}{8pt}
    \begin{tabular}{ccccccccc}
    \hline
    Models&Supernova mechanisms&$k$&$N_{\rm tot}$&$N_{\rm 15}$&$N_{\rm mg}$&$N_{\rm 15}/N_{\rm tot}$&$N_{\rm mg}/N_{\rm tot}$&$N_{\rm mg}/N_{15}$\\
    \hline
    R0&rapid&0&$4.27\times 10^2$&$4.27\times 10^2$&0&100\%&0\%&0\%\\
    R3&rapid&3&$6.95\times 10^2$&$6.53\times 10^2$&0&94\%&0\%&0\%\\
    \hline
    D0&delayed&0&$1.28\times 10^4$&$1.28\times 10^4$&$1.16\times 10^4$&100\%&90\%&90\%\\
    D3&delayed&3&$1.42\times 10^4$&$1.27\times 10^4$&$4.17\times 10^3$&89\%&29\%&33\%\\
    \hline
    S0&stochastic&0&$2.86\times 10^3$&$2.86\times 10^3$&$6.85\times 10^2$&100\%&24\%&24\%\\
    S3&stochastic&3&$1.06\times 10^4$&$6.83\times 10^3$&$4.48\times 10^2$&65\%&4\%&7\%\\
    \hline
    \end{tabular}
    \label{tab:res}
\end{table*}

We then use a Monte Carlo method to simulate the evolution of a large population of primordial binaries. The binary evolution code \texttt{BSE} \citep{Hurley+2002+BSE,Kiel+2006,Shao+2014+Be} is utilized to model the formation and evolution of incipient BH LMXBs. The lookback time $t_{\rm lb}$ for the birth of primordial binaries is assumed to follow a uniform distribution within the past $12~{\rm Gyr}$ and a constant star formation rate $3~{\rm M_\odot~yr^{-1}}$ is applied. The initial mass $M_{1,0}$ of the primary is sampled from a power-law distribution $p(M_{1,0})\propto M_{1,0}^{-2.7}$ \citep{Kroupa+1993} in the range of $5-100~{\rm M_\odot}$. We assume that the mass ratio $q=M_{2,0}/M_{1,0}$ between the secondary ($M_{2,0}$) and the primary masses has a uniform distribution between $0$ and $1$, and the initial orbital separation $A_0$ follows $p(A_0)\propto A_0^{-1}$ in the range of $3~{\rm R_\odot}-10^4~{\rm R_\odot}$. We take the eccentricity of all the primordial binaries to be zero, as its effect on the formation and evolution of LMXBs is minor \citep{Hurley+2002+BSE}. We use the energy conversation equation \citep{Webbink+1984} to treat the CEE, taking the CEE efficiency $\alpha_{\rm CE}=1.0$. We use the numerically calculated results by  \cite{Xu+2010} and \cite{Wang+2016} for the binding energy parameter $\lambda_{\rm CE}$ of the stellar envelope for a variety of massive and intermediate-mass stars.

To deal with the compact remnant masses and possible natal kicks, we adopt the rapid \citep{Fryer+2012}, delayed \citep{Fryer+2012} and stochastic supernova scenarios \citep{Mandel+2020}, which are widely used in the literature. Note that the rapid supernova mechanism predicts an inherent mass gap around $2.5-5~{\rm M_\odot}$. For the first two mechanisms, the natal kick $v_{\rm k}$ of BHs are assumed to obey a Maxwellian distribution with the dispersion $\sigma_{\rm k}=265~{\rm km~s^{-1}}$ \citep{Hobbs+2005}  reduced by a material fallback factor $f_{\rm fb}$ \citep[for more details, see][]{Shao+2021}. For the stochastic supernova mechanism, we use the natal kick proposed by \citet{Mandel+2020}. The minimum mass of BHs at birth is assumed to be $2.5~{\rm M_\odot}$. We construct six models R0, R3, D0, D3, S0, and S3, corresponding to different supernova prescriptions and different values of $k$. The parameters and results of these six models are presented in \autoref{tab:res}.

\section{Results and comparison with observations}\label{sec:res}

\begin{figure*}[!ht]
    \centering
    \includegraphics[width=\linewidth]{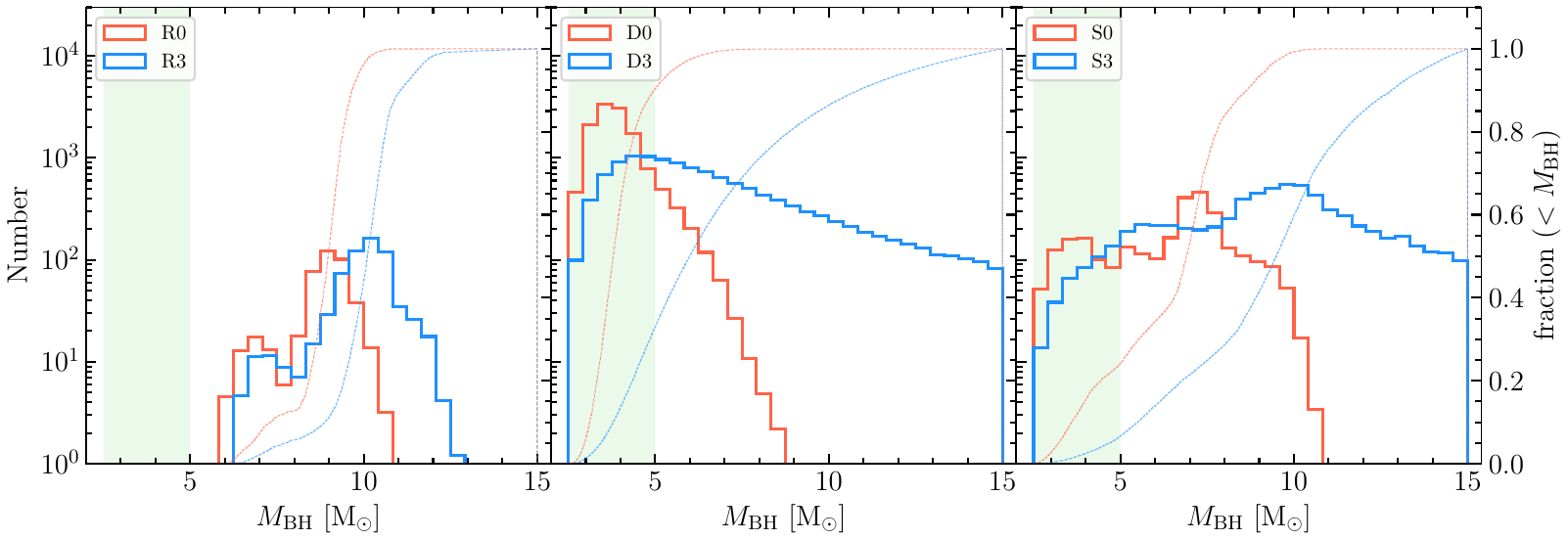}
    \caption{Mass distribution of BHs in I/LMXBs. The left, middle and right panels show the results of the rapid, delayed and stochastic supernova models. The red and blue solid histograms correspond to $k=0$ and $k=3$ cases, respectively. The red and blue dashed lines show the normalized cumulative mass distribution in the range of $2.5-15~{\rm M_\odot}$ in $k=0$ and $k=3$ cases, respectively. The shaded region indicates the mass gap ($2.5-5~{\rm M_\odot}$).\label{fig:MbhXrb}}
\end{figure*}

\begin{figure*}[!ht]
    \centering
    \includegraphics[width=\linewidth]{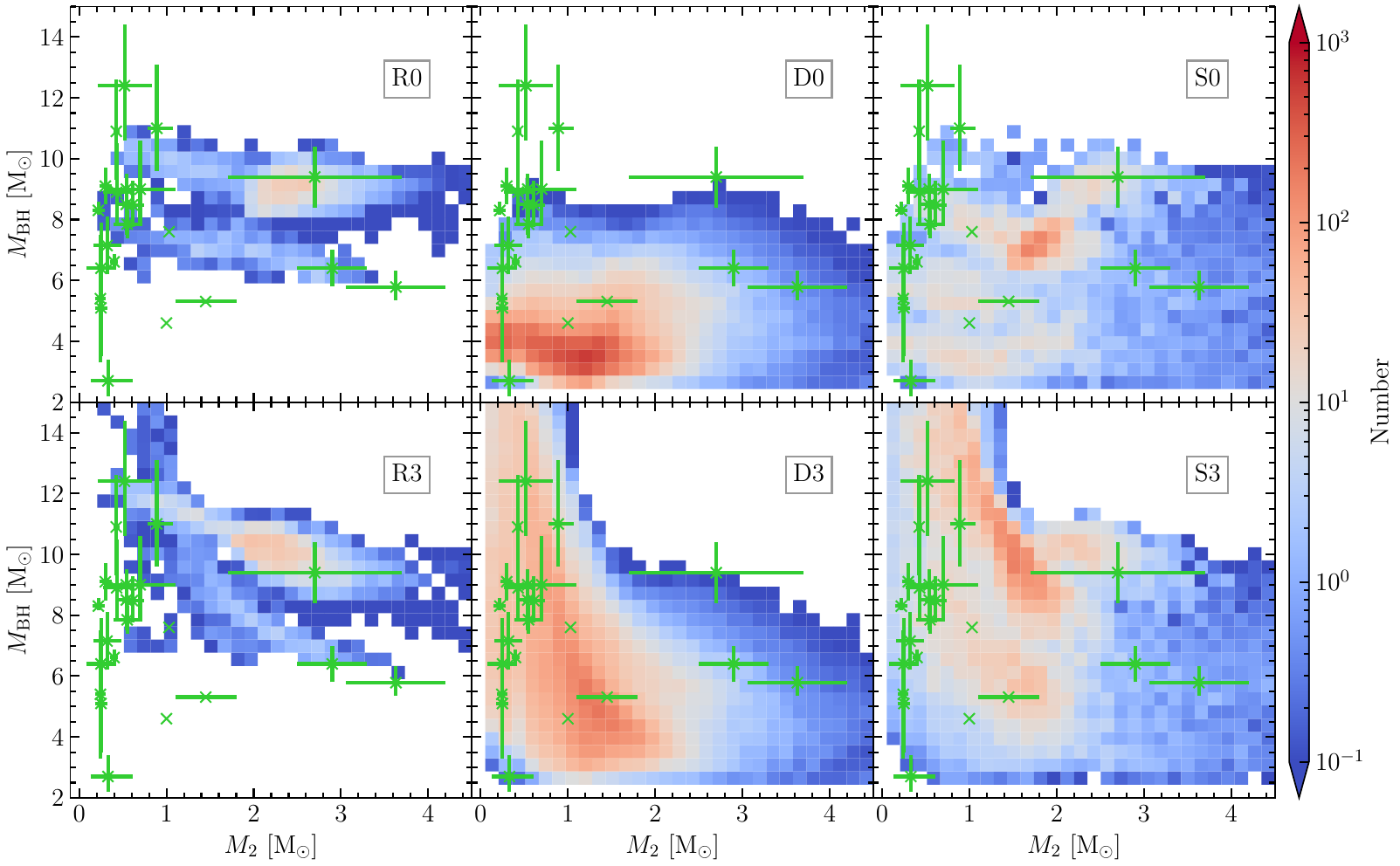}
    \caption{Number distribution of I/LMXBs in the donor mass-BH mass ($M_2$--$M_{\rm BH}$) plane. The upper and lower panels correspond $k=0$ and $k=3$ cases, respectively. The left, middle and right panels represent the results in the rapid, delayed and stochastic supernova models, respectively. The number of I/LMXBs in each pixel is represented by different colors. The green crosses with errors show the observed L/IMXBs (see \autoref{tab:obs}).\label{fig:m2mbh}}
\end{figure*}

\begin{figure*}[!ht]
    \centering
    \includegraphics[width=\linewidth]{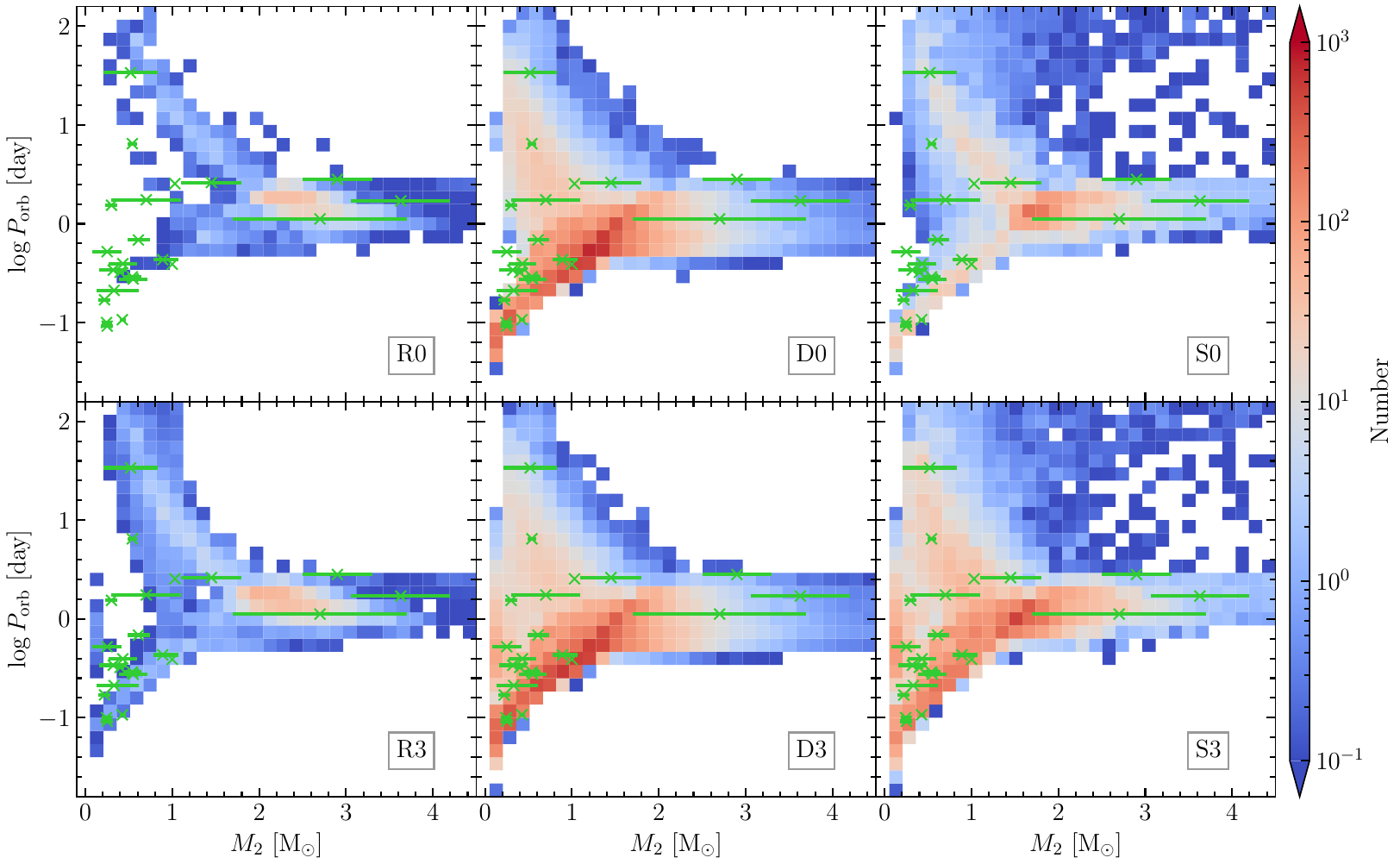}
    \caption{Similar to \autoref{fig:m2mbh} but the donor mass-orbital period ($M_2-P_{\rm orb}$) distributions.
    \label{fig:m2porb}}
\end{figure*}

We select BH binaries with a $0.08~{\rm M_\odot}<M_2<6~{\rm M_\odot}$ non-degenerate donor star  and BH mass accretion rates $>10^{-12}~{\rm M_\odot~yr^{-1}}$ as I/LMXBs. Their predicted numbers in different BH mass range are presented in \autoref{tab:res}. \autoref{fig:MbhXrb} shows the mass function of BHs. The left, middle and right panels depict the results of Models R0/R3, D0/D3 and S0/S3, respectively. The red and blue solid histograms demonstrate the number distribution of the BH mass in the case of $k=0$ and $k=3$ in the range of $2.5-15~{\rm M_\odot}$, respectively. The red ($k=0$) and blue ($k=3$) dashed lines show the normalized cumulative mass distribution, and the shaded region indicates the mass gap ($2.5-5~{\rm M_\odot}$). In the traditional formation models (R0, D0, and S0), because the envelope of stars of mass $>45\,\rm M_\sun$ is too heavy to be expelled by a low-mass companion during CEE, the masses of BHs are confined to be $\sim 9-11~{\rm M_\odot}$, but if considering cosmologically-coupled mass growth  (Models~R3, D3, and S3), the final masses of BHs can reach several hundred ${\rm M_\odot}$, so we artificially cut off at $15~{\rm M_\odot}$ in the figure. We find that BHs with mass $<15~{\rm M_\odot}$ constitute $\sim 65\%-94\%$ of the total BH I/LMXB population in this case (see \autoref{tab:res}).

While Model R0 produces no BHs with mass $\lesssim 5~{\rm M_\odot}$,  Models~D0 and S0 predict that $\sim 90\%$ and $\sim 24\%$ of the BHs are mass-gap BHs, respectively. In comparison, only $\sim 33\%$ and $\sim 7\%$ of BHs with masses $<15~{\rm M_\odot}$ fall within the mass gap in Models~D3 and S3, respectively.

\autoref{fig:m2mbh} shows the number distribution of I/LMXBs in the donor mass - BH mass ($M_2$--$M_{\rm BH}$) plane. The upper and lower panels correspond $k=0$ and $k=3$ cases, respectively. The left, middle and right panels show the results of Models R0/R3, D0/D3 and S0/S3, respectively. The number of I/LMXBs in each pixel is reflected by different colors. The green crosses represent the observed I/LMXBs (see \autoref{tab:obs}). It is seen that, none of the three supernova scenarios in the case of $k=0$ can well reproduce the observed I/LMXBs -- in Model R0, the BH masses range from $\sim 6$ to $\sim 11~{\rm M_\odot}$, and in Models~D0 an S0, the BH masses are less than $\sim 9$ and $\sim 10~{\rm M_\odot}$, respectively. The situation is improved in the $k=3$ case, with the help of the cosmologically-coupled mass growth. Especially most of the observed BHs can be covered in Models~D3 and S3. However, the relative numbers in different $M_2$--$M_{\rm BH}$ regions still seem not precisely match the observations. In particular, the predicted donor masses in Models D3 and S3 are considerably larger than observations. Since low-mass stars are unlikely to survive CEE during the first mass transfer process, this implies that either abnormally high values of
the CE efficiency is required or the majority of BHs may originate from failed supernovae \citep[for discussion, see][and references therein]{Li+2015}.

\autoref{fig:m2porb} shows the number distribution in the donor mass - orbital period ($M_2-P_{\rm orb}$) plane. 
Both Models~R0 and R3 are hard to reproduce the observed BH LMXBs, while Models D and S seem to work better, no matter $k=0$ or 3. The cosmologically-coupled mass growth of BHs causes the orbit to shrink and RLOF earlier, leading to a higher birth rate for BH I/LMXBs. In addition, the duration of the mass transfer is prolonged (see \autoref{tab:mesaRes} and the upper panels of \autoref{fig:mesa}). So,
more compact BH LMXBs can be produced in Models D3 and S3 compared with Models D0 and S0. This seems to be compatible with the observed distribution.

\section{Discussion and conclusions} \label{sec:dis}
In this work, we employ both binary evolution calculation and binary population synthesis to simulate the formation and evolution of LMXBs. The goal of this work is to examine the possible effect of  cosmologically-coupled mass growth of BHs, especially on the origin of the mass gap. Our main results are summarized as follows.

(1) The cosmologically-coupled mass growth of BHs seems to naturally result in the mass gap or dearth around $2.5-5\,\rm M_\odot$. The percentage of mass gap BHs decreases from $90\%$ and $24\%$ in Models D0 and S0 to $29\%$ and $4\%$ in Models D3 and S3.

(2) The cosmologically-coupled mass growth of BHs always leads to orbital contraction. This can trigger RLOF in originally wide binaries, producing more compact LMXBs, which are difficult to form in the standard CEE theory, because a low-mass companion star does not have enough orbital energy to expel the envelope of the BH's progenitor star.

(3) The cosmologically-coupled mass growth extends the BH masses up to several hundred solar mass. Future observations of such intermediate-mass BH binaries will be crucial in testing the feasibility of the model.

There are some discussions on how the cosmologically-coupled mass growth of BHs can be constrained by observations in the literature.
Gaia~BH1 \citep{El-Badry+2023} and BH2 \citep{El-Badry+2023b} are two detached BH binaries with the BH masses being $9.62~{\rm M_\odot}$ and $8.9~{\rm M_\odot}$, respectively. The orbital periods are $186~{\rm day}$ and $1277~{\rm day}$, and the eccentricities are $0.45$ and $0.52$ for Gaia~BH1 and BH2, respectively. In the framework of isolated binary evolution, \cite{Andrae+2023} suggested that, if the BHs are cosmologically coupled to the expansion of the Universe, there is a $70\%-77\%$  probability that the initial BH masses are below $2.2~{\rm M_\odot}$. However, the current binary orbits can not accommodate the BH progenitors in the supergiant phase, and the low-mass optical companions are unable to eject the massive envelope during CEE. Thus, the two BHs seem unlikely to origin from isolated binary evolution \citep[as discussed in][]{El-Badry+2023,El-Badry+2023b}. 
However, if BHs are coupled to the expansion of the Universe, it is possible that these two BH binaries formed from extremely wide binaries. The initial low-mass BHs grew with the expansion of the Universe, resulting in a contraction of the orbit while conserving the eccentricity imparted by the supernova kick.

\cite{Rodriguez+2023} argued that if $k=3$, two BH candidates  in the globular cluster NGC~3201 \citep[][]{Giesers+2018,Giesers+2019} must have near face-on orientations or have formed from stellar collapse with NS-like masses ($<2.2~{\rm M_\odot}$). While the latter is not impossible, the cosmologically-coupled mass growth may shift low-mass BHs to be more massive and leave the mass gap. Alternatively, low-mass BHs may be formed from accretion induced collapse of NSs \citep[e.g.,][]{Gao+2022}. In this situation, the cosmological coupling may not be significant.

The evolutionary tracks in \autoref{sec:method} show that, after the mass transfer, the growth of the BHs, together with gravitational radiation,  continues to drive the orbital contraction, and the final orbital periods can be as short as a few hours within the Hubble time. During this stage the binaries will manifest as low-frequency gravitational wave sources. The forthcoming low-frequency gravitational wave detectors \citep[e.g., LISA,][]{LISA} could detect these binaries and test whether the BH evolution is influenced by the expansion of the Universe. Considering the cosmologically-coupled mass growth of BHs (assuming $k=3$), \cite{Ghodla+2023} showed that the estimated rates of gravitational-wave merger events would be three orders of magnitude higher than the observed rates. The LIGO-Virgo-KAGRA collaboration is currently conducting the O4 observing run. It is anticipated that a larger sample of BH mergers at different redshift will provide valuable test of the cosmological coupling of BHs.

\vspace{5mm}

\begin{acknowledgements}
We thank Yuan-Xing, Gao (高原兴); Lei, Feng (冯磊); Yi-Ying, Wang (王艺颖) and Lei, Lei (雷磊) for helpful discussions. We are also grateful to an anonymous referee for useful comments and suggestions. This work was supported by the National Key Research and Development Program of China (2021YFA0718500), the Natural Science Foundation of China under grant No.~12041301, 12121003, and Project~U1838201 supported by NSFC and CAS.
\end{acknowledgements}

\facilities{The computation was made by using the facilities at the High-Performance Computing Center of Collaborative Innovation Center of Advanced Microstructures (Nanjing University, \url{https://hpc.nju.edu.cn}).}

\software{\texttt{MESA} \citep{MESA+2011,MESA+2013,MESA+2015,MESA+2018,MESA+2019,MESA+2023}, \texttt{BSE} \citep{Hurley+2002+BSE}, \texttt{Numpy} \citep{numpy}, \texttt{Astorpy} \citep{astropy}, \texttt{Matplotlib} \citep{matplotlib}, \texttt{Jupyter-lab} (\url{https://jupyterlab.readthedocs.io}).}

\bibliography{reference.bib}{}
\bibliographystyle{aasjournal}

\end{CJK*}
\end{document}